\documentclass[12pt,preprint]{aastex}
\slugcomment{Accepted to Astronomical Journal}

\shorttitle{NIR Variability of S5 0716$+$714}
\shortauthors{Gupta et al.}

\begin{document}

\title{Multi-color Near Infra-red Intra-day and Short Term Variability \\
of the Blazar S5 0716$+$714}

\author{Alok C. Gupta\altaffilmark{1}, Sang-Mok Cha\altaffilmark{2},
Sungho Lee\altaffilmark{2}, Ho Jin\altaffilmark{2}, Soojong Pak\altaffilmark{3},
Seoung-hyun Cho\altaffilmark{2}, Bongkon Moon\altaffilmark{2},
Youngsik Park\altaffilmark{2}, In-Soo Yuk\altaffilmark{2},
Uk-won Nam\altaffilmark{2}, and Jaemann Kyeong\altaffilmark{2}}
\altaffiltext{1}{Aryabhatta Research Institute of Observational Sciences (ARIES),
Manora Peak, \\
\hspace*{0.22in} Nainital $-$ 263129, India.}
\altaffiltext{2}{Korea Astronomy and Space Science Institute
(KASI), Daejeon, South Korea.}
\altaffiltext{3}{Kyung Hee University, Department of Astronomy and Space Science,
South Korea.}

\email{leesh@kasi.re.kr, Phone No. +82-42-865-3354, Fax No. +82-42-865-3272}

\begin{abstract}
In this paper, we report results of our near-infrared (NIR) photometric variability
studies of the BL Lacertae object S5 0716$+$714. NIR photometric observations spread
over 7 nights during our observing run April 2$-$9, 2007 at 1.8 meter telescope
equipped with KASINICS (Korea Astronomy and Space Science Institute Near Infrared
Camera System) and J, H, and Ks filters at Bohyunsan Optical Astronomy Observatory (BOAO),
South Korea. We searched for intra-day variability, short term variability and color
variability in the BL Lac object. We have not detected any genuine intra-day variability
in any of J, H, and Ks passbands in our observing run. Significant short term variability
$\sim$ 32.6\%, 20.5\% and 18.2\% have been detected in J, H, Ks passbands, respectively,
and $\sim$ 11.9\% in (J-H) color.
\end{abstract}


\section{Introduction}

Blazars constitute a small subclass of the most enigmatic class of radio-loud active
galactic nuclei (AGNs) consisting of BL Lacertae objects (BL Lacs) and flat spectrum
radio quasars (FSRQs). BL Lacs show largely featureless optical continuum. Blazars
exhibit strong flux variability at all wavelengths of the complete electromagnetic
(EM) spectrum, strong polarization ($> 3\%$) from radio to optical wavelengths,
usually core dominated radio structures, and predominantly nonthermal radiation at
all wavelengths. In a unified model of radio-loud AGNs based on the angle between
the line of sight and the emitted jet from the source, blazars jets make angle of
$<$ 10$^{\circ}$ from the line of sight (Urry \& Padovani 1995).

From the study of the spectral energy distributions (SEDs) of blazars, it is found
that blazars SEDs have two peaks (Fossati et al. 1998; Ghisellini et al. 1998).
The first component peaks at near-infrared (NIR)/optical in low energy peaked blazars
(LBLs) and at UV/X-rays in high energy peaked blazars (HBLs). The second component
peaks at GeV energies in LBLs and at TeV energies in HBLs. The EM emission is dominated
by synchrotron component at low energies and at high energies probably by the inverse
Compton (IC) component (Coppi 1999, Sikora \& Madejski 2001, Krawczynski 2004).

From observations of blazars, it is known that they vary on the diverse time scales
ranging from a few minutes to several years. Blazars variability can be broadly divided
into three classes viz. intra-day variability (IDV) or intra-night variability or
micro-variability, short term variability and long term variability. Variations in
flux of a few tenth of a magnitude over the course of a day or less is often known
as IDV (Wagner \& Witzel 1995). Short and long term variabilities can have time
scales from a few weeks to several months and several months to years, respectively.

First convincing optical IDV in blazar was reported by Miller et al. (1989), and since
then variability of blazars on diverse time scales in radio to optical bands have been
studied extensively and results have been reported in large number of papers (e.g. Carini
1990; Mead et al. 1990; Takalo et al. 1992, 1996; Heidt \& Wagner 1996;
Sillanp\"a\"a et al. 1996a, 1996b; Bai et al. 1998, 1999; Fan et al. 1998,
2002, 2007; Xie, et al. 2002; Ciprini et al. 2003, 2007; Gupta et al. 2004, 2008a,
2008b; Stalin et al. 2005 and references therein). In a recent paper, Gupta \& Joshi
(2005) have done statistical analysis of occurrence of optical IDV in different classes
of AGNs. They divided their sample of 113 optical lights curves of blazars in three
different time durations and found 64\%(18/28), 63\%(29/46), and 82\%(32/39) blazars show
IDV if observed for $\leq$3h, 3h to $\leq$6h, and $>$6h, respectively.

S5 0716$+$714 ($\alpha_{2000.0} =$ 07:21:53.4, $\delta_{2000.0} = +$71:20:36.4) is one of
the brightest BL Lac object which has featureless optical continuum. The non detection of
its host galaxy first sets a lower limit of redshift z $>$ 0.3 (Wagner et al. 1996),
and then z $>$ 0.52 (Sbarufatti et al. 2005). Very recently, Nilsson et al. (2008) have
claimed of its host galaxy detection which produces a ``standard candle'' value of
$z = 0.31 \pm 0.08$. Wagner \& Witzel (1995) reported that the duty cycle of the source
is one which implies that the source is almost always in the
active state. The variability of S5 0716$+$714 has been studied in the complete EM spectrum on
all time scales (e.g. Raiteri et al. 2003; Gupta et al. 2008a and references therein).
Large and variable optical polarization in the source has been reported (Takalo et al. 1994;
Fan et al. 1997; Impey et al. 2000). Since 1994, this source has been extensively
monitored in optical bands. There are 5 major optical outbursts reported in the source:
at the beginning of 1995, in late 1997, in the fall of 2001, in March 2004
and in the beginning of 2007 (Raiteri et al. 2003; Foschini et al. 2006; Gupta et al.
2008a). These 5 outbursts give possible period of long term variability
of $\sim$ 3.0$\pm$0.3 years.

Compared to radio and optical bands, there are only a few attempts to search for NIR flux
variability on diverse timescales in blazars (e.g. Mead et al. 1990; Takalo et al. 1992;
Gupta et al. 2004; Hagen-Thorn et al. 2006 and references therein). Since blazars emit
radiations in the complete EM spectrum, they are ideal candidates for multi-wavelength
observations. But either due to unavailability of good quality NIR detectors or unavailability
of low humidity observing sites at several 1-2 meter class NIR/optical telescopes around the
world, there were no focused effort to search for NIR flux variability in LBLs in which SEDs
synchrotron component peaks in NIR/optical bands. Now we have an excellent opportunity to
carry out such observations from the Bohyunsan Optical Astronomy Observatory (BOAO), South
Korea which has 1.8 meter telescope and is equipped with KASINICS (Korea Astronomy and Space
Science Institute Near Infrared Camera System) (Moon et al. 2008). We have recently started our long term pilot
project to search for flux variability on diverse time scales in LBLs. Our present and future
planned observations will fill the gap between radio and optical bands and will give deep insight
into the important and less studied NIR flux variability properties of LBLs. Simultaneous radio
to gamma-ray observations will be useful to detect the synchrotron and IC component peaks
of LBLs from the SEDs and will be useful to understand the emission mechanism of LBLs in the
complete EM spectrum.
With this motivation, we recently carried out J, H, Ks bands photometric
observations of our first target, the BL Lac object S5 0716$+$714 which is a LBL, in
7 observing nights during April 2 $-$ April 9, 2007.

The paper is structured as follows: section 2 describes observations and data reduction
methods, in section 3 we mention our results, the discussions and conclusions of the present
work is reported in section 4.

\section{Observations and Data Reduction}

The time-series observations of S5 0716+714 were carried out in J, H and Ks bands
during April 2$-$9, 2007 using KASINICS mounted at the 1.8 m telescope of BOAO
in South Korea. On April 4, observations were
made only in J and Ks bands. The on-source integration time in each band was 120 sec
throughout the observations except 60 sec on April 4. The KASINICS has field of view of
$\sim$ $3.3 \times 3.3$ arcmin$^2$ with a 512 $\times$ 512 InSb array ALADDIN III
Quadrant. Image frames were obtained in four dithered positions, offset by
$\approx$ 15 arcsec. In all image frames of S5 0716$+$714, two standard stars,
Star no. 3 and Star no. 5 of Villata et al. (1998) were always present.

Each image frame was processed to subtract the dark, to correct
the pixel-to-pixel inhomogeneity (flat fielding), to remove the bad pixels and cosmic
rays. In order to remove the sky background, we subtracted another dithered image at
a different position from an object image frame. Then four dithered images were
aligned and average-combined to improve the signal-to-noise (S/N) ratio.
Instrumental magnitudes of the standard stars and
the blazar in the processed image frames were measured by the aperture photometry.
Data reductions and deriving the instrumental magnitudes were performed using
standard routines in IRAF\footnote{IRAF is distributed by the National Optical Astronomy Observatories,
which are operated by the Association of Universities for Research in Astronomy, Inc.,
under cooperative agreement with the National Science Foundation.} (Image Reduction
and Analysis Facility) software. Additional C programs were developed locally for
automated data processing.

Since the object and the standard stars were observed
in the same image frame, no correction for atmospheric extinction was done. Two
standard stars in the blazar field were used to check the non-variable characteristics
of standard stars and finally one standard star (Star no. 3) was used to calibrate the
instrumental magnitudes of the blazar S5 0716+714. The reliability of the photometry is
verified by differential magnitudes of standard stars (Star no. 5 $-$ Star no. 3).
The results remain consistent throughout the observations within 1 sigma of 0.021, 0.028
and 0.028 in J, H and Ks bands, respectively.

\section{Results}

\subsection{Variability Detection Criterion}

Time variability of the blazar S5 0716$+$714 is investigated by using the
parameter C
(Romero et al. 1999) defined as the average of C$_{1}$ and C$_{2}$
\begin{eqnarray}
C_{1} = \frac {\sigma (BL - Star A)}{\sigma (Star A - Star B)} \hspace*{0.2in} {\rm{and}} \hspace*{0.2in}
C_{2} = \frac {\sigma (BL - Star B)}{\sigma (Star A - Star B)}
\end{eqnarray}
Using aperture photometry of the blazar and two standard stars in the blazar field, we
determined the differential instrumental magnitude of the blazar $-$ standard star A,
blazar $-$ standard star B and standard star A $-$ standard star B. We determined observational
scatter from blazar $-$ standard star A ($\sigma$ (BL $-$ Star A)), blazar $-$ standard star B
($\sigma$ (BL $-$ Star B)) and standard star A $-$ standard star B ($\sigma$ (Star A $-$ Star B)).

If C $>$ 2.57, the confidence limit of variability is 99\%.
The typical uncertainty level in our calculation of C parameter itself is 10$-$20\%.
Here we used Star no. 5 and
3 of Villata et al. (1998) as Star A and Star B respectively. J, H, Ks magnitude of these
standard stars are taken from the 2MASS catalogue\footnote{www.ipac.caltech.edu/2mass/}
(Skrutskie et al. 2006). Final calibration of S5 0716$+$714 data is done by Star B (Star no. 3).
Photometric software in IRAF does not give actual internal error of brightness but it
gives photon noise. The internal photometric errors of brightness for each J, H and K$_s$ band
are estimated using artificial add star experiment as described by Stetson (1987).
We found that the standard deviation ($\sigma$) in each J, H and K$_s$ band
is $\sim$ 1.5 times larger than the typical photon noise.
So, the typical photometric error in each J, H and K$_s$ band is $\sim$ 0.01 mag.

\subsection{Intra-Day Variability}

\noindent
{\bf J Passband}

We observed the blazar S5 0716$+$714 on April 2, 3, 4, 6, 7, 8 and 9, 2007 in J passband.
The light curves of the blazar (filled circles) and differential instrumental magnitude
(star 5 - star 3) (filled triangles) with different arbitrary offsets are displayed in
different panels in the left column of the Fig. 1. Date of observations are marked in the
panels. We performed the IDV detection test described above by equation (1), and got the
value of C for April 2, 3, 4, 6, 7, 8 and 9, 2007 are 0.50, 0.75, 0.64, 0.50, 1.04, 1.83
and 2.30, respectively (see Table 1), which confirms that the source has not shown genuine IDV on any
night of our 7 nights of observations in J passband. Photometric data of the observing
campaign in J passband is reported in Table 2.

\noindent
{\bf H Passband}

We observed the blazar S5 0716$+$714 on April 2, 3, 6, 7, 8 and 9, 2007 in H passband.
The light curves of the blazar (filled circles) and differential instrumental magnitude
(star 5 - star 3) (filled triangles) with different arbitrary offsets are displayed in
different panels in the middle column of the Fig. 1. Date of observations are marked in
the panels. We performed the IDV detection test described above by equation (1), and got
the value of C for April 2, 3, 6, 7, 8 and 9, 2007 are 0.49, 0.79, 2.05, 0.62, 1.22 and
1.32, respectively (Table 1), which confirms that the source has not shown genuine IDV on any night
of our 6 nights of observations in H passband. Photometric data of the observing campaign
in H passband is reported in Table 3.

\noindent
{\bf Ks Passband}

We observed the blazar S5 0716$+$714 on April 2, 3, 4, 6, 7, 8 and 9, 2007 in Ks passband.
The light curves of the blazar (filled circles) and differential instrumental magnitude
(star 5 - star 3) (filled triangles) with different arbitrary offsets are displayed in
different panels in the right column of the Fig. 1. Date of observations are marked in
the panels. We performed the IDV detection test described above by equation (1), and got
the value of C for April 2, 3, 4, 6, 7, 8 and 9, 2007 are 7.00, 0.54, 0.63, 1.35, 0.78,
1.20 and 1.37, respectively (Table 1), which confirms that the source has not shown genuine IDV
on April 3, 4, 6, 7, 8 and 9, 2007 in Ks passband.
The value of C on April 2, 2007 show the source has shown IDV but data points
are only 2, so, its reliability is doubtful. Photometric data of the observing campaign
in Ks passband is reported in Table 4.

\subsection{Short Term Variability}

In Fig. 2, nightly averaged light curves of S5 0716$+$714 (standard mag.) and
comparison stars (differential instrumental mag. of Stars 3 and 5) in J, H, Ks,
J$-$H and H$-$Ks are plotted in the different panels from bottom to top, respectively.
Here we estimate the 99\% confidence detection level of short term variability
using the variability detection test described in section 3.1.
For a specific date of observations, we have
taken mean time of all the image frames in J , H and Ks bands and then converted the
mean time in Julian date (JD).

Short term variability amplitude is calculated by using the following relation
(Heidt \& Wagner 1996)
\begin{eqnarray}
A = 100 \times \sqrt {(A_{max} - A_{min})^{2} - 2\sigma^{2}} \hspace*{0.1in}\%
\end{eqnarray}
where A$_{max}$ and A$_{min}$ are the maximum and minimum magnitude in the calibrated
light curve of the blazar in complete observing run. $\sigma$ is the averaged measurement
error of the blazar light curve.
Errors in our determination of A are less than 1\%.

\noindent
{\bf J Passband}

The short term light curve of S5 0716$+$714 in J passband is displayed in the bottom
panel of Fig. 2. The maximum variation noticed in the source is 0.325 mag (between
its faintest level at 12.488 mag on JD 2454200.11894 and the brightest level at 12.163
mag on JD 2454194.08202). The value of C is calculated to be 7.05,
which supports the existence of short term variation
in the source in J band observations. We calculated short term variability amplitude using
equation (2) and found that the source has varied $\sim$ 32.6\%.

\noindent
{\bf H Passband}

The short term light curve of S5 0716$+$714 in H passband is displayed in the second
panel from bottom in Fig. 2. The maximum variation noticed in the source is 0.206 mag
(between its faintest level at 11.731 mag on JD 2454200.11894 and the brightest level
at 11.525 mag on JD 2454194.08202). The parameter C is 4.14, which supports the existence of short
term variation in the source in H band observations. We calculated short term variability
amplitude using equation (2) and found that source has varied $\sim$ 20.5\%.

\noindent
{\bf Ks Passband}

The short term light curve of S5 0716$+$714 in Ks passband is displayed in the third
panel from bottom in Fig. 2. The maximum variation noticed in the source is 0.183 mag
(between its faintest level at 10.936 mag on JD 2454200.11894 and the brightest level
at 10.753 mag on JD 2454194.08202). The parameter C is 3.93, which supports the existence of short
term variation in the source in Ks band observations. We calculated short term
variability amplitude using equation (2) and found that source has varied $\sim$ 18.2\%.

\noindent
{\bf J$-$H Color}

The short term light curve of S5 0716$+$714 in J$-$H color is displayed in the second
panel from top in Fig. 2. The maximum variation noticed in the source is 0.119 mag
(between level at 0.757 mag on JD 2454200.11894 and at the level at 0.638 mag on
JD 2454194.08202). The parameter C is 2.86, which supports the existence of short
term variation in J$-$H color. We calculated short term
variability amplitude using equation (2) and found that source has varied $\sim$ 11.9\%.

\noindent
{\bf H$-$Ks Color}

The short term light curve of S5 0716$+$714 in H$-$Ks color is displayed in the top
panel in Fig. 2. The maximum variation noticed in the source is 0.061 mag (between level
at 0.795 mag on JD 2454200.11894 and at the level at 0.734 mag on JD 2454197.14597).
The parameter C is 1.78. Therefore, no H$-$Ks color variation in the source is detected
in our observations.

\subsection{Spectral Energy Distribution}

Spectral behavior of the blazar S5 0716$+$714 in different nights of our observations
are displayed by different symbols in Fig. 3. S5 0716$+$714 is a LBL in which synchrotron
component peaks in NIR/optical bands. The spectral behavior of the source in our 7 nights
observations indicates that the synchrotron component would peak at wavelengths
shorter than our J band.
Since there is no simultaneous data in radio and optical bands,
we cannot determine the peak position.

The source was brighter on April 3 compared to April 2, then it became fainter day after
day until our last observation on April 9.
The SED data of each day is fitted by a power-law function and we got the spectral index
of $-0.87$, $-0.76$, $-0.86$, $-0.88$, $-0.93$, and $-1.00$ for April 2, 3, 6, 7, 8 and 9,
respectively. The spectral index increases before the maximum brightness on April 3 and
decreases after it down to $-1$. The changes in spectral index between the maximum ($-0.76$)
and the minimum ($-1.00$) is 0.24. This SED variation is owing to the larger
flux variation at
shorter wavelengths and is consistent with the short-term variability amplitudes
reported in section 3.3.

\section{Discussions and Conclusions}

From our multi-band NIR observations of the blazar S5 0716$+$714 in 7 observing nights
in April 2007, genuine IDV in any of J, H and Ks is not detected. We noticed the existence
of significant short term flux variability in the blazar from our observations. The total
short term variation detected in our observations in J, H, Ks passbands are $\sim$ 32.6\%,
20.5\% and 18.2\%, respectively. Our data show significant variation in J$-$H color
($\sim$ 11.9\%) but H$-$Ks color variation was not detected. The difference in
short term variations in H and Ks passbands is only 2.3\% which caused no genuine H$-$Ks
color variation. We have noticed variable spectral index (ranging from $-0.76$ to $-1.00$)
with a mean value of $\sim -0.88$ in our observations. The variable spectral index is mainly
due to variation in J band flux. We also found correlated flux and spectral index (higher
the flux, higher the spectral index).
We observed the source in the post outburst
state. Outburst of the source was reported by Gupta et al. (2008a) in their January
$-$ February 2007 observations and they also noticed in their March 2007 observations,
the source was becoming fainter than January $-$ February 2007.

S5 0716$+$714 was the target of three simultaneous multi-wavelength campaigns
(Tagliaferri et al. 2003; Ostorero et al. 2006; Villata et al. 2008) and also several
monitoring campaigns in single or two EM bands (e.g. radio-optical, optical and optical-X-ray)
(Wagner et al. 1990; Quirrenbach et al. 1991; Sagar et al. 1999; Villata et al. 2000; Foschini
et al. 2006; and references therein). It has shown radio and optical IDVs during all radio to
optical campaigns (Heeschen et al. 1987; Wagner et al. 1990, 1996; Ghisellini et al. 1997;
Sagar et al. 1999; Quirrenbach et al. 2000; Raiteri et al. 2003; Agudo et al. 2006; Gupta et al. 2008a,
and references therein). It is the first IDV source in which simultaneous variations in radio and
optical bands were detected which indicated a possible intrinsic origin of the observed IDV
(Wagner et al. 1990; Quirrenbach et al. 1991). VLBI observations of the source over more than
20 years, show a very compact source at centimeter wavelengths with an evidence of a core-dominated
jet extending several tens of milliarcseconds to the North (Eckart et al. 1986, 1987; Witzel et al.
1988; Polatidis et al. 1995; Jorstad et al. 2001).
The X-ray observations have shown strong variations with short flares ($\approx$ 1000s) detected
with the {\it ROentgen SATellite} (ROSAT) (Cappi et al. 1994). The source was also detected in
hard X-rays up to 60 keV when observed after the outburst state of 2000 (Tagliaferri et al. 2003).
It has been detected by {\it Energetic Gamma-Ray Experiment Telescope} (EGRET) onboard the
{\it Compton Gamma-Ray Observatory} CGRO at GeV energies with steep $\gamma-$ray spectrum
(Hartman et al. 1999). But the soft $\gamma-$ray part of its SED is poorly known and upper limit
of the source detection in 3$-$10 MeV energy from the {\it imaging COMPton TELescope} (COMPTEL)
was reported by Collmar (2006). An exceptional energy sampling data of the blazar
was obtained in the simultaneous multi-wavelength observing campaign in November 2003
(Ostorero et al. 2006). The source was very bright at radio frequencies and in a rather low
optical state (R = 14.17 - 13.64). Significant short term variability and IDV were detected
in the radio bands.
The source was not detected by INTEGRAL in the observing campaign but the upper limit of the
source emission in 3$-$200 keV was estimated. On September 2007, the source was
detected in $\gamma-$rays by the recently launched satellite {\it Astro-rivelatore Gamma
a Immagini LEggero} (AGILE) (Villata et al. 2008).

Several models have been developed to explain the IDV and short term variability in
radio-loud AGNs viz. the shock-in-jet models, accretion disk based models (e.g. Wagner
\& Witzel 1995; Urry \& Padovani 1995; Ulrich et al. 1997; and references therein).
For blazars in the outburst state, IDV and short term variability are
strongly supported by the jet based models of radio-loud AGNs. In general, blazars
emission in the outburst state is nonthermal Doppler boosted emission from jets (Blandford
\& Rees 1978; Marscher \& Gear 1985; Marscher et al. 1992; Hughes et al. 1992). On the
other hand, IDV and short term variability of blazars in the low-state can be explained
by the models based on some kind of instability in the accretion disk (e.g. Mangalam \&
Wiita 1993; Chakrabarti \& Wiita 1993).

Here we rule out the possibility of emission from the accretion disk because it is expected
to be relevant only in the source low-state
(the source was in the post outburst period).
The source was in the outburst state $\sim$ 2 months before the present observations
(Gupta et al. 2008a). In the low-state, jet emission is less dominant over the thermal
emission from the accretion disk. According to the unified scheme of radio-loud AGNs,
blazars are seen nearly face on, so, any fluctuations on the accretion disk should
produce a detectable change in the emission characteristics.

The observed short term variability in the blazar S5 0716$+$714 is
possibly explained by the jet based model known as turbulent jet model. According to
this model, Marscher et al. (1992) suggested that the Reynolds number in the relativistic
jet should be very high which will cause turbulent jet plasma. The shock will impinge upon
regions of slightly different magnetic field strengths, densities and velocities, so the
observed flux is expected to vary. The timescales of this proposed type of variations
are shorter and their amplitudes larger at higher frequencies. In the observations
reported here, we got variability amplitudes of $\sim$ 32.6\%, 20.5\% and 18.2\% in
J, H, Ks bands, respectively, implying larger variations at higher frequencies.
The relation between flux variation amplitude and frequency is confirmed
from the SED and color variation.

Since the duty cycle of the source is 1, we may expect to detect IDV. However, we have
not detected any genuine IDV in our observations. It might be due to less numbers of
data points and short durations of the observing runs during each night, hence
in the future we plan to observe the source for a longer time each night.
This will allow us to collect more data points and possibly to have a higher
signal-to-noise in order to have higher sensitivity to possible genuine IDVs.

\acknowledgments

We thank the referee for useful comments. We thank Yoon Ho Park and the staff at BOAO
for their excellent support during our observations.

\clearpage

\begin{deluxetable}{cccrrrcccl}
\tabletypesize{\scriptsize}
\tablecaption{Results of intra-day variability of the blazar S5 0716$+$714$^{*}$.
\label{tbl-1}}
\tablewidth{0pt}
\tablehead{
\colhead{Date} & \colhead{Band} & \colhead{N} & \colhead{Diff. mag}     & \colhead{Diff. mag}
& \colhead{Diff. mag}         & \colhead{Variable} & \colhead{C} \\
\colhead{dd.mm.yyyy} &                &             & \colhead{BL - S$_{A}$} & \colhead{BL - S$_{B}$}
& \colhead{S$_{A}$ - S$_{B}$} &          &
}
\startdata
02.04.2007 & J     & ~2 & 0.950$\pm$0.028 & $-$0.092$\pm$0.005 & 1.042$\pm$0.033 & NV & 0.50  \\
           & H     & ~2 & 0.486$\pm$0.023 & $-$0.513$\pm$0.025 & 0.999$\pm$0.049 & NV & 0.49  \\
           & Ks    & ~2 & $-$0.181$\pm$0.006 & $-$1.182$\pm$0.008 & 1.001$\pm$0.001 & PV & 7.00  \\
           & J$-$H & ~2 & 0.464$\pm$0.036 & 0.421$\pm$0.025    & 0.043$\pm$0.059 & NV & 0.52   \\
           & H$-$Ks& ~2 & 0.667$\pm$0.024 & 0.669$\pm$0.026    & $-$0.002$\pm$0.049 & NV & 0.51  \\
03.04.2007 & J   & 13 & 0.842$\pm$0.014 & $-$0.164$\pm$0.016   & 1.007$\pm$0.020 & NV & 0.75  \\
           & H   & 13 & 0.462$\pm$0.028 & $-$0.524$\pm$0.026   & 0.985$\pm$0.034 & NV & 0.79  \\
           & Ks  & 13 & $-$0.225$\pm$0.006 & $-$1.245$\pm$0.021 & 1.020$\pm$0.025 & NV & 0.54  \\
           & J$-$H & 13 & 0.380$\pm$0.031 & 0.360$\pm$0.031     & 0.022$\pm$0.039 & NV & 0.79   \\
           & H$-$Ks& 13 & 0.687$\pm$0.029 & 0.721$\pm$0.033     & $-$0.035$\pm$0.042 & NV & 0.74  \\
04.04.2007 & J   & 16 & 0.928$\pm$0.008 & $-$0.079$\pm$0.015    & 1.007$\pm$0.018 & NV & 0.64   \\
           & Ks  & 16 & $-$0.195$\pm$0.009 & $-$1.191$\pm$0.020 & 0.996$\pm$0.023 & NV & 0.63   \\
06.04.2007 & J   & ~2 & 0.973$\pm$0.013 & $-$0.045$\pm$0.004    & 1.017$\pm$0.017 & NV & 0.50   \\
           & H   & ~2 & 0.494$\pm$0.028 & $-$0.492$\pm$0.017    & 0.986$\pm$0.011 & NV & 2.05  \\
           & Ks  & ~2 & $-$0.156$\pm$0.056 & $-$1.165$\pm$0.122 & 1.009$\pm$0.066 & NV & 1.35   \\
           & J$-$H & ~2 & 0.479$\pm$0.031 & 0.447$\pm$0.017     & 0.031$\pm$0.020 & NV & 1.20   \\
           & H$-$Ks& ~2 & 0.650$\pm$0.063 & 0.673$\pm$0.123     & $-$0.023$\pm$0.067 & NV & 1.39  \\
07.04.2007 & J   & ~9 & 1.044$\pm$0.014 & $-$0.001$\pm$0.011    & 1.044$\pm$0.012 & NV & 1.04   \\
           & H   & ~9 & 0.595$\pm$0.012 & $-$0.438$\pm$0.009    & 1.033$\pm$0.017 & NV & 0.62  \\
           & Ks  & ~9 & $-$0.097$\pm$0.008 & $-$1.130$\pm$0.020 & 1.033$\pm$0.018 & NV & 0.78   \\
           & J$-$H & ~9 & 0.449$\pm$0.018 & 0.437$\pm$0.014     & 0.011$\pm$0.021 & NV & 0.76   \\
           & H$-$Ks& ~9 & 0.692$\pm$0.014 & 0.692$\pm$0.022     & 0.000$\pm$0.025 & NV & 0.72  \\
08.04.2007 & J   & 14 & 1.094$\pm$0.025 &    0.077$\pm$0.041    & 1.017$\pm$0.018 & NV & 1.83   \\
           & H   & 14 & 0.625$\pm$0.021 & $-$0.391$\pm$0.023    & 1.016$\pm$0.018 & NV & 1.22   \\
           & Ks  & 14 & $-$0.083$\pm$0.021 & $-$1.104$\pm$0.046 & 1.021$\pm$0.028 & NV & 1.20   \\
           & J$-$H & 14 & 0.469$\pm$0.033 & 0.468$\pm$0.047     & 0.001$\pm$0.025 & NV & 1.60   \\
           & H$-$Ks& 14 & 0.708$\pm$0.030 & 0.713$\pm$0.051     & $-$0.005$\pm$0.033 & NV & 1.23  \\
09.04.2007 & J   & ~9 & 1.169$\pm$0.024 &    0.142$\pm$0.022    & 1.027$\pm$0.010 & NV & 2.30   \\
           & H   & ~9 & 0.668$\pm$0.018 & $-$0.347$\pm$0.011    & 1.015$\pm$0.011 & NV & 1.32   \\
           & Ks  & ~9 & $-$0.042$\pm$0.016 & $-$1.080$\pm$0.010 & 1.038$\pm$0.019 & NV & 1.37   \\
           & J$-$H & ~9 & 0.501$\pm$0.030 & 0.489$\pm$0.025     & 0.012$\pm$0.015 & NV & 1.83   \\
           & H$-$Ks& ~9 & 0.710$\pm$0.024 & 0.733$\pm$0.015     & $-$0.023$\pm$0.022 & NV & 0.89  \\
\enddata

\tablenotetext{*} {V, PV and NV in the Variable column represent variable, possible variable and non variable
respectively. N represents the number of data points.}

\end{deluxetable}

\clearpage

\begin{deluxetable}{cccc}
\tabletypesize{\scriptsize}
\tablecaption{J band Photometric Data of the blazar S5 0716$+$714.
\label{tbl-2}}
\tablewidth{0pt}
\tablehead{
\colhead{Date} & \colhead{UT} & \colhead{Magnitude}     & \colhead{Error} \\
\colhead{yyyy.mm.dd} & \colhead{hour}  &             &
}
\startdata
2007.04.02  &  13.0767 &  12.249 & 0.0103    \\
2007.04.02  &  13.6577 &  12.289 & 0.0108   \\
2007.04.03  &  11.3628 &  12.165 & 0.0067  \\
2007.04.03  &  11.6856 &  12.168 & 0.0058 \\
2007.04.03  &  12.1344 &  12.190 & 0.0058\\
2007.04.03  &  12.4653 &  12.172 & 0.0058  \\
2007.04.03  &  13.1033 &  12.152 & 0.0067   \\
2007.04.03  &  13.5543 &  12.160 & 0.0067  \\
2007.04.03  &  14.0155 &  12.153 & 0.0067 \\
2007.04.03  &  14.3276 &  12.164 & 0.0067   \\
2007.04.03  &  14.7036 &  12.162 & 0.0072   \\
2007.04.03  &  15.1420 &  12.172 & 0.0081  \\
2007.04.03  &  15.5005 &  12.130 & 0.0081    \\
2007.04.03  &  15.8084 &  12.163 & 0.0081   \\
2007.04.03  &  16.1480 &  12.146 & 0.0081  \\
2007.04.04  &  13.5451 &  12.232 & 0.0058    \\
2007.04.04  &  13.6240 &  12.241 & 0.0058   \\
2007.04.04  &  13.6928 &  12.239 & 0.0058  \\
2007.04.04  &  13.7610 &  12.247 & 0.0058    \\
2007.04.04  &  13.8381 &  12.240 & 0.0058   \\
2007.04.04  &  13.9185 &  12.245 & 0.0058  \\
2007.04.04  &  13.9874 &  12.240 & 0.0058    \\
2007.04.04  &  14.0717 &  12.238 & 0.0067   \\
2007.04.04  &  14.1499 &  12.255 & 0.0058  \\
2007.04.04  &  14.2285 &  12.261 & 0.0058    \\
2007.04.04  &  14.3353 &  12.255 & 0.0067   \\
2007.04.04  &  14.4575 &  12.248 & 0.0081  \\
2007.04.04  &  14.5378 &  12.252 & 0.0067    \\
2007.04.04  &  14.6199 &  12.258 & 0.0067   \\
2007.04.04  &  14.6928 &  12.254 & 0.0072  \\
2007.04.04  &  14.7733 &  12.245 & 0.0067    \\
2007.04.06  &  15.1859 &  12.301 & 0.0058   \\
2007.04.06  &  15.5757 &  12.282 & 0.0081  \\
2007.04.07  &  13.2219 &  12.337 & 0.0058    \\
2007.04.07  &  13.4691 &  12.355 & 0.0058   \\
2007.04.07  &  13.7392 &  12.370 & 0.0058  \\
2007.04.07  &  14.0084 &  12.384 & 0.0067    \\
2007.04.07  &  14.2890 &  12.373 & 0.0058   \\
2007.04.07  &  14.5922 &  12.364 & 0.0067  \\
2007.04.07  &  14.9806 &  12.347 & 0.0058    \\
2007.04.07  &  15.2281 &  12.370 & 0.0067   \\
2007.04.07  &  15.4945 &  12.367 & 0.0067  \\
2007.04.08  &  11.1906 &  12.455 & 0.0076    \\
2007.04.08  &  11.4216 &  12.455 & 0.0076   \\
2007.04.08  &  11.6621 &  12.452 & 0.0076  \\
2007.04.08  &  11.9226 &  12.423 & 0.0067    \\
2007.04.08  &  12.1747 &  12.417 & 0.0067   \\
2007.04.08  &  13.4435 &  12.392 & 0.0067  \\
2007.04.08  &  13.6623 &  12.403 & 0.0076    \\
2007.04.08  &  13.9838 &  12.413 & 0.0067   \\
2007.04.08  &  14.2744 &  12.388 & 0.0076  \\
2007.04.08  &  14.5426 &  12.416 & 0.0076    \\
2007.04.08  &  15.0147 &  12.397 & 0.0058   \\
2007.04.08  &  15.3563 &  12.401 & 0.0058  \\
2007.04.08  &  15.6473 &  12.379 & 0.0058    \\
2007.04.08  &  15.9339 &  12.392 & 0.0067   \\
2007.04.09  &  13.5311 &  12.440 & 0.0076  \\
2007.04.09  &  13.9564 &  12.483 & 0.0076    \\
2007.04.09  &  14.2050 &  12.473 & 0.0058   \\
2007.04.09  &  14.4568 &  12.490 & 0.0058  \\
2007.04.09  &  14.7123 &  12.503 & 0.0058    \\
2007.04.09  &  14.9869 &  12.507 & 0.0058   \\
2007.04.09  &  15.3351 &  12.498 & 0.0058  \\
2007.04.09  &  15.6168 &  12.524 & 0.0067    \\
2007.04.09  &  15.9539 &  12.478 & 0.0058   \\
\enddata

\end{deluxetable}

\clearpage

\begin{deluxetable}{cccc}
\tabletypesize{\scriptsize}
\tablecaption{H band Photometric Data of the blazar S5 0716$+$714.
\label{tbl-3}}
\tablewidth{0pt}
\tablehead{
\colhead{Date} & \colhead{UT} & \colhead{Magnitude}     & \colhead{Error} \\
\colhead{yyyy.mm.dd} & \colhead{hour}  &             &
}
\startdata
2007.04.02  &  13.1916 &  11.532 & 0.0072   \\
2007.04.02  &  13.7532 &  11.565 & 0.0081   \\
2007.04.03  &  11.4691 &  11.537 & 0.0050  \\
2007.04.03  &  11.8890 &  11.528 & 0.0042 \\
2007.04.03  &  12.2291 &  11.528 & 0.0050\\
2007.04.03  &  12.6306 &  11.554 & 0.0042   \\
2007.04.03  &  13.3047 &  11.478 & 0.0050   \\
2007.04.03  &  13.6283 &  11.483 & 0.0050  \\
2007.04.03  &  14.1066 &  11.521 & 0.0050 \\
2007.04.03  &  14.4783 &  11.530 & 0.0050\\
2007.04.03  &  14.8032 &  11.533 & 0.0057   \\
2007.04.03  &  15.2618 &  11.516 & 0.0064   \\
2007.04.03  &  15.6085 &  11.510 & 0.0072  \\
2007.04.03  &  15.9071 &  11.589 & 0.0071 \\
2007.04.03  &  16.2718 &  11.514 & 0.0058    \\
2007.04.06  &  15.3187 &  11.577 & 0.0057   \\
2007.04.06  &  15.7229 &  11.537 & 0.0092  \\
2007.04.07  &  13.3143 &  11.642 & 0.0050 \\
2007.04.07  &  13.5536 &  11.638 & 0.0042    \\
2007.04.07  &  13.8537 &  11.669 & 0.0050   \\
2007.04.07  &  14.1191 &  11.651 & 0.0036  \\
2007.04.07  &  14.3754 &  11.663 & 0.0050 \\
2007.04.07  &  14.7183 &  11.671 & 0.0050    \\
2007.04.07  &  15.0560 &  11.661 & 0.0036   \\
2007.04.07  &  15.3214 &  11.663 & 0.0036  \\
2007.04.07  &  15.5968 &  11.664 & 0.0050 \\
2007.04.08  &  11.2673 &  11.714 & 0.0050    \\
2007.04.08  &  11.4962 &  11.728 & 0.0050   \\
2007.04.08  &  11.7304 &  11.715 & 0.0050  \\
2007.04.08  &  12.0194 &  11.706 & 0.0050 \\
2007.04.08  &  13.2942 &  11.687 & 0.0050    \\
2007.04.08  &  13.5133 &  11.683 & 0.0050   \\
2007.04.08  &  13.7582 &  11.677 & 0.0050  \\
2007.04.08  &  14.0690 &  11.671 & 0.0050 \\
2007.04.08  &  14.3667 &  11.652 & 0.0050    \\
2007.04.08  &  14.6240 &  11.685 & 0.0180   \\
2007.04.08  &  15.1095 &  11.664 & 0.0036  \\
2007.04.08  &  15.4417 &  11.681 & 0.0042 \\
2007.04.08  &  15.7355 &  11.674 & 0.0036    \\
2007.04.08  &  16.0395 &  11.688 & 0.0050   \\
2007.04.09  &  13.7980 &  11.718 & 0.0036  \\
2007.04.09  &  14.0295 &  11.717 & 0.0050 \\
2007.04.09  &  14.2900 &  11.723 & 0.0050\\
2007.04.09  &  14.5404 &  11.735 & 0.0050   \\
2007.04.09  &  14.7849 &  11.711 & 0.0045   \\
2007.04.09  &  15.0437 &  11.737 & 0.0050  \\
2007.04.09  &  15.4440 &  11.723 & 0.0045 \\
2007.04.09  &  15.7708 &  11.757 & 0.0036\\
2007.04.09  &  16.0364 &  11.761 & 0.0042   \\
\enddata

\end{deluxetable}

\clearpage

\begin{deluxetable}{cccc}
\tabletypesize{\scriptsize}
\tablecaption{Ks band Photometric Data of the blazar S5 0716$+$714.
\label{tbl-4}}
\tablewidth{0pt}
\tablehead{
\colhead{Date} & \colhead{UT} & \colhead{Magnitude}     & \colhead{Error} \\
\colhead{yyyy.mm.dd} & \colhead{hour}  &             &
}
\startdata
2007.04.02  &  13.5410 &  10.793 & 0.0064  \\
2007.04.02  &  13.8485 &  10.802 & 0.0064  \\
2007.04.03  &  11.5766 &  10.755 & 0.0042\\
2007.04.03  &  12.0182 &  10.774 & 0.0042 \\
2007.04.03  &  12.3193 &  10.766 & 0.0042  \\
2007.04.03  &  12.9633 &  10.765 & 0.0042   \\
2007.04.03  &  13.4103 &  10.768 & 0.0042 \\
2007.04.03  &  13.7719 &  10.729 & 0.0050  \\
2007.04.03  &  14.1940 &  10.742 & 0.0050   \\
2007.04.03  &  14.5875 &  10.743 & 0.0050 \\
2007.04.03  &  14.9135 &  10.754 & 0.0050  \\
2007.04.03  &  15.3751 &  10.764 & 0.0057   \\
2007.04.03  &  15.7039 &  10.773 & 0.0057 \\
2007.04.03  &  16.0015 &  10.740 & 0.0064  \\
2007.04.03  &  16.3985 &  10.716 & 0.0064   \\
2007.04.04  &  13.5776 &  10.786 & 0.0050 \\
2007.04.04  &  13.6598 &  10.785 & 0.0050  \\
2007.04.04  &  13.7267 &  10.783 & 0.0050   \\
2007.04.04  &  13.7985 &  10.787 & 0.0050 \\
2007.04.04  &  13.8766 &  10.793 & 0.0050  \\
2007.04.04  &  13.9520 &  10.785 & 0.0050  \\
2007.04.04  &  14.0338 &  10.776 & 0.0050\\
2007.04.04  &  14.1124 &  10.793 & 0.0057 \\
2007.04.04  &  14.1902 &  10.774 & 0.0050  \\
2007.04.04  &  14.2609 &  10.777 & 0.0050  \\
2007.04.04  &  14.4067 &  10.777 & 0.0064\\
2007.04.04  &  14.5010 &  10.767 & 0.0057 \\
2007.04.04  &  14.5757 &  10.780 & 0.0057  \\
2007.04.04  &  14.6575 &  10.791 & 0.0057  \\
2007.04.04  &  14.7303 &  10.802 & 0.0057\\
2007.04.04  &  14.8092 &  10.776 & 0.0057 \\
2007.04.06  &  15.4085 &  10.783 & 0.0050  \\
2007.04.06  &  15.8120 &  10.862 & 0.0170  \\
2007.04.07  &  13.3938 &  10.897 & 0.0050\\
2007.04.07  &  13.6301 &  10.877 & 0.0050 \\
2007.04.07  &  13.9250 &  10.872 & 0.0064  \\
2007.04.07  &  14.1989 &  10.885 & 0.0057  \\
2007.04.07  &  14.4659 &  10.878 & 0.0057\\
2007.04.07  &  14.9096 &  10.873 & 0.0050 \\
2007.04.07  &  15.1434 &  10.882 & 0.0050  \\
2007.04.07  &  15.4132 &  10.884 & 0.0057  \\
2007.04.07  &  15.6956 &  10.884 & 0.0057\\
2007.04.08  &  11.3472 &  10.943 & 0.0057 \\
2007.04.08  &  11.5714 &  10.901 & 0.0057  \\
2007.04.08  &  11.8080 &  10.907 & 0.0057  \\
2007.04.08  &  12.0997 &  10.909 & 0.0057\\
2007.04.08  &  13.3733 &  10.902 & 0.0057 \\
2007.04.08  &  13.5831 &  10.902 & 0.0057  \\
2007.04.08  &  13.8364 &  10.902 & 0.0057  \\
2007.04.08  &  14.1698 &  10.896 & 0.0057\\
2007.04.08  &  14.4618 &  10.874 & 0.0057 \\
2007.04.08  &  13.9064 &  10.908 & 0.0057  \\
2007.04.08  &  15.2049 &  10.872 & 0.0050  \\
2007.04.08  &  15.5335 &  10.862 & 0.0050\\
2007.04.08  &  15.8278 &  10.881 & 0.0050 \\
2007.04.08  &  16.1308 &  10.870 & 0.0050  \\
2007.04.09  &  13.8834 &  10.917 & 0.0057  \\
2007.04.09  &  14.1093 &  10.917 & 0.0057\\
2007.04.09  &  14.3602 &  10.922 & 0.0050 \\
2007.04.09  &  14.6315 &  10.933 & 0.0050  \\
2007.04.09  &  14.8720 &  10.932 & 0.0042  \\
2007.04.09  &  15.2103 &  10.960 & 0.0057\\
2007.04.09  &  15.5332 &  10.955 & 0.0057 \\
2007.04.09  &  15.8579 &  10.951 & 0.0042  \\
2007.04.09  &  16.1326 &  10.939 & 0.0042   \\
\enddata

\end{deluxetable}

\clearpage

\begin{figure}
\plotone{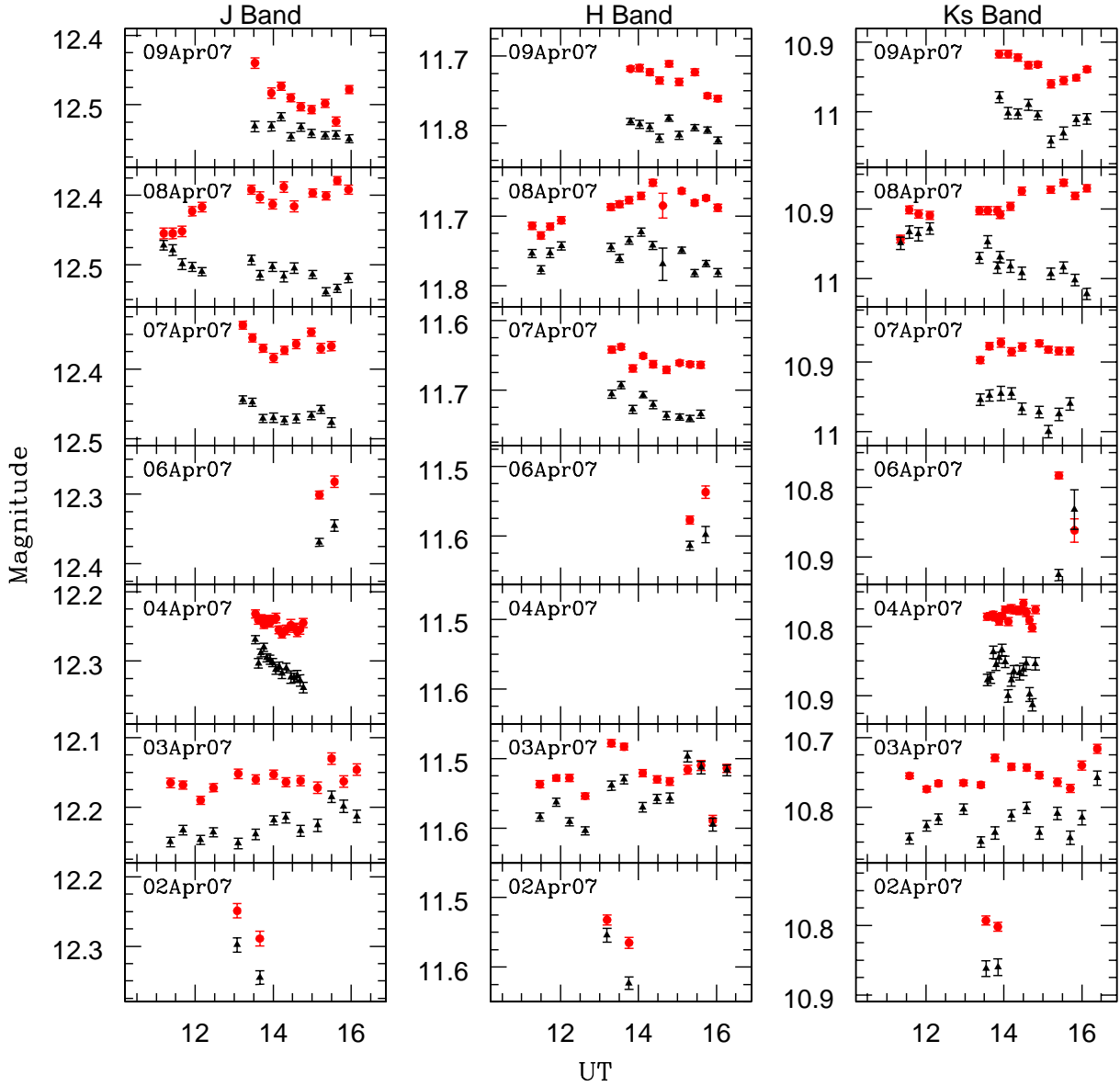}
\caption{The J, H, Ks bands light curves of S5 0716$+$718 (filled circles) and differential
instrumental magnitude of standard stars (Star5$-$Star3) (filled triangles) during the nights
from April 2, 2007 to April 9, 2007. Panels in left, middle and right columns show the light
curves in J, H, Ks bands respectively. Standard stars differential light curve is offset
for clarity by different arbitrary constants on the all 7 nights of observations.}
\end{figure}

\begin{figure}
\plotone{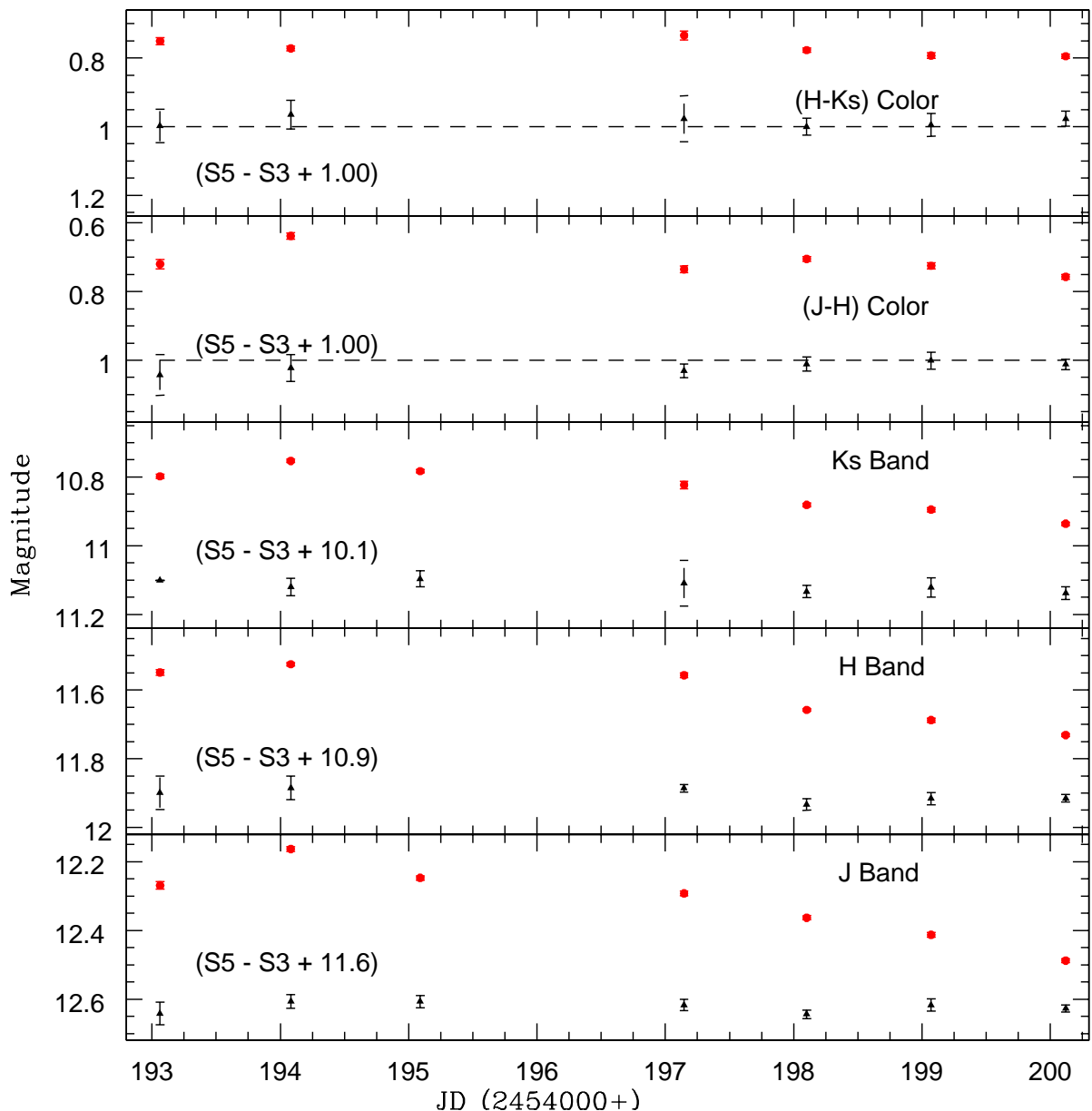}
\caption{The daily averaged light curves in J, H, Ks, J$-$H and H$-$Ks of S5 0716$+$714 (filled
circles) and differential instrumental magnitude of standard stars (Star5$-$Star3) (filled
triangles) during the nights of April 2, 2007 - April 9, 2007 are plotted in the different
panels from bottom to top respectively. For clarity, the differential instrumental magnitude
of standard stars (Star5$-$Star 3) are offset by the amounts marked on the panels.}
\end{figure}

\begin{figure}
\plotone{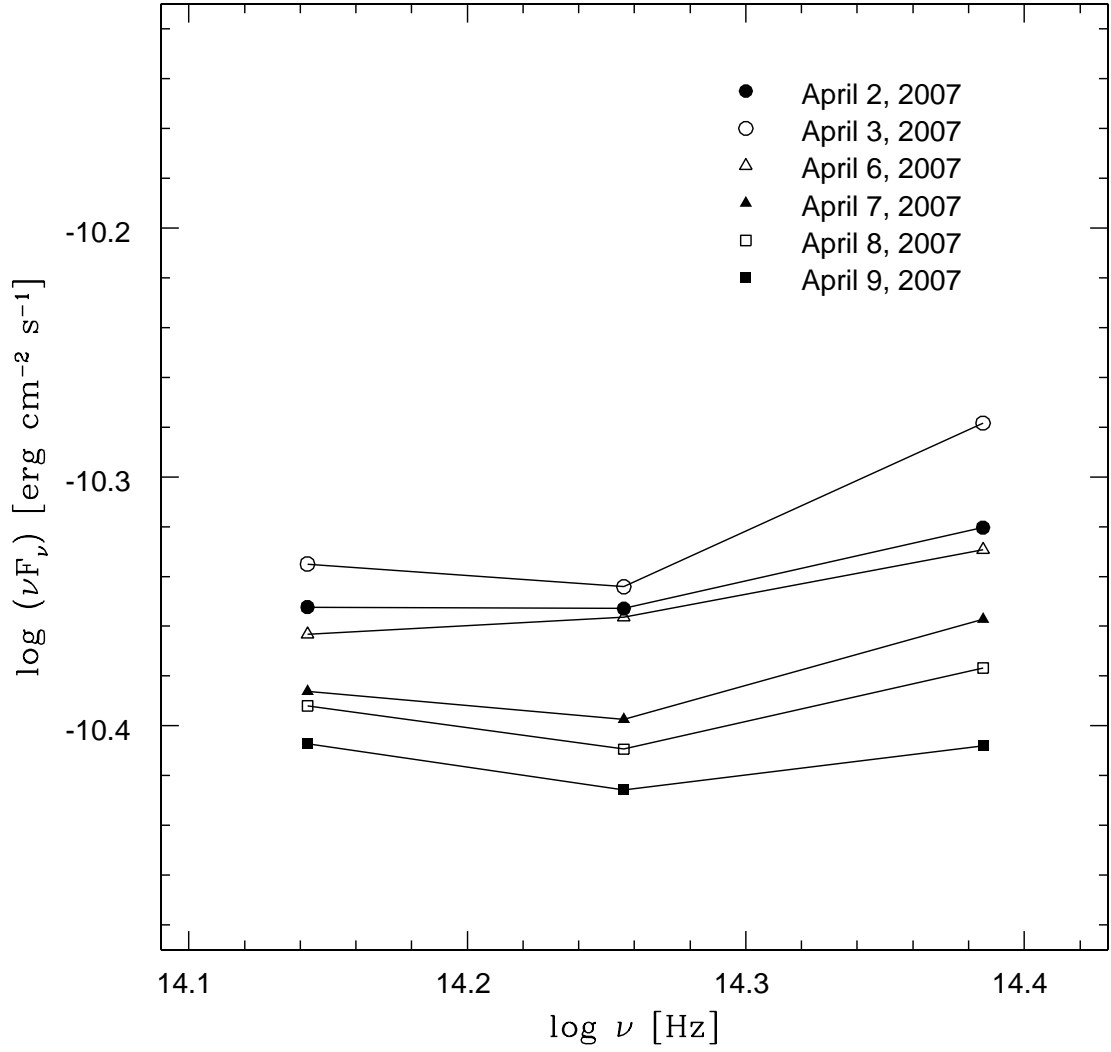}
\caption{Spectral behavior of S5 0716$+$714 for 6 nights during April 2, 2007 - April 9, 2007.
Different symbols are used for different dates.}
\end{figure}

\end{document}